\documentclass[11pt]{article}  
\usepackage{graphicx}  
\usepackage{amssymb,amsmath}   

\begin{document}

\title{Over-populated gauge fields on the lattice}
\author{J{\"u}rgen Berges, S{\"o}ren Schlichting, D\'enes Sexty\\
Universit\"at Heidelberg, Institut f{\"u}r Theoretische Physik,\\
Philosophenweg 16, 69120~Heidelberg, Germany}

\date{}

\maketitle

\begin{abstract}
We study nonequilibrium dynamics of $SU(2)$ pure gauge theory starting from initial over-population, where intense classical gauge fields are characterized by a single momentum scale $Q_s$. Classical-statistical lattice simulations indicate a quick evolution towards an approximate scaling behavior with exponent $3/2$ at intermediate times. Remarkably, the value for the scaling exponent may be understood as arising from the leading ${\cal O}(g^2)$ contribution in the presence of a time-dependent background field. The phenomenon is associated to weak wave turbulence describing an energy cascade towards higher momenta. This particular aspect is very similar to what is observed for scalar theories, where an effective cubic interaction arises because of the presence of a time-dependent Bose condensate.
\end{abstract}

\section{Motivation and overview}

The question of thermalization is one of
the most pressing issues in the physics of ultra-relativistic heavy-ion collisions. These experiments involve far-from-equilibrium
systems of strongly interacting matter described by quantum chromodynamics
(QCD). The data reveal robust collective phenomena, whose theoretical understanding from QCD  represents a great challenge \cite{EXP}.
This concerns even the theoretically ``clean case'' of very large
nuclei and energies. In this idealized limit, the
matter formed shortly after the collision may be described
by classical gluon fields with a characteristic
``saturation'' momentum scale $Q_s$ that grows with the energy and the
size of nuclei. Since $Q_s$ is large, the strong coupling constant $\alpha(Q_s)$ is
small. However, the gluons at saturation are strongly correlated because
their low momentum modes have very high occupancy $\sim 1/\alpha$. 

Recently, it has been argued that in the earlier stages of a heavy-ion collision at sufficiently high energies, the parametrically large gluon density may lead to the phenomenon of Bose condensation~\cite{Blaizot:2011xf}. This has to be compared to alternative scenarios \cite{Kurkela:2011ti}. For a recent review see Ref.~\cite{EMMI}. Inelastic, particle number changing processes preclude the possibility that the true equilibrium state be a Bose condensate, but there is the possibility that a transient condensate develops during the evolution of the system. Finally, this question is related to the competing time scales for elastic as compared to inelastic processes in gauge theories. However, the weak-coupling analysis is complicated by the fact that modes with occupancies $\sim 1/\alpha$ require non-perturbative descriptions. 

Important aspects of this non-perturbative gluon dynamics can be described 
by classical-statistical lattice gauge theory simulation 
techniques \cite{Romatschke:2005pm,Berges:2007re,Kunihiro:2010tg,Fukushima:2011nq}.
Such classical field simulations, with suitable
averages over the initial conditions, can
accurately describe the dynamics if the expectation values of field anti-commutators are much larger than the corresponding commutators \cite{Berges:2004yj}. Loosely speaking, this is realized in the presence of sufficiently high occupation numbers per mode, $n(p)$. This concerns regimes 
of large occupation ($n(p) \gg 1$) all
the way up to the over-populated regime, where $n(p)\sim 1/\alpha$. They
fail of course when typical occupations become smaller than unity, at which
point quantum effects need to be taken into account. The range of validity of classical-statistical simulations has been tested to high accuracy for self-interacting scalar field
theories~\cite{Aarts:2001yn,Arrizabalaga:2004iw,Berges:2008wm} and 
theories with fermions~\cite{Berges:2010zv}, where appropriate
far-from-equilibrium approximation techniques are available also directly for the
respective quantum field theory. 

In particular, the dynamical formation of a 
Bose condensate was recently demonstrated for scalar quantum fields starting from initial over-population \cite{Berges:2012us}. It has been shown that Bose condensation occurs as a consequence of an inverse particle cascade with a universal power-law spectrum. This particle transport towards low momenta is part of a dual cascade, in which energy is also transfered by weak wave turbulence towards higher momenta. Exponents associated with
these two cascades are under analytical control and are well
reproduced in the simulations \cite{Berges:2008wm,Berges:2008sr}. 
In particular, the value of the
exponent for the UV cascade can be understood as arising from the
existence of the condensate leading to an effective cubic interaction \cite{Micha:2004bv}.
Condensation was also discussed in Ref.~\cite{Dusling:2010rm}, and similar dynamics 
was analyzed in Ref.~\cite{Nowak:2010tm} in the context of cold atoms emphasizing also the role of non-trivial topological configurations~\cite{Gasenzer:2011by}. Whereas a Bose condensate
develops and remains present in the relativistic system until eventually inelastic
scattering depletes it, for the non-relativistic theory with conserved particle number no decay of the condensate due to number changing processes is observed \cite{Berges:2012us}.

In this work, we study the non-abelian dynamics of over-populated gauge fields for the $SU(2)$ gauge group. A major focus is to work out further the relevant differences and similarities between gauge and scalar degrees of freedom out of equilibrium in this context. The most remarkable result will be that classical simulations starting from initial over-population indicate for Yang-Mills theory the same turbulence exponent as for scalars. This appears non-trivial for the following reasons. In scalar theories, the fact that the interactions are
nearly local in momentum space typically plays a crucial role to
explain the flow of momenta. This is apparently not the case in gauge
theories, where one may go from hard to soft momenta in one
collision \cite{Mueller:2006up}. 

Furthermore, for scalar theories the inelastic processes are of higher order in the coupling than elastic collisions. Therefore, one expects that the role of number changing processes, which can compete with condensate formation, can be suppressed for weak coupling in scalar theories. In contrast, elastic and inelastic processes are parametrically of the same order in the gauge theory. It is, therefore, a question beyond simple parametric estimates whether condensation occurs from initial over-population or not. 

We explain that both scalar and gauge theories can show the same turbulent scaling exponents in the presence of a time-dependent background field. This can be done using resummed perturbation theory, since the phenomenon of weak wave turbulence occurs in the kinetic regime with occupancies in the range $1/\alpha \gg n(p) \gg 1$ \cite{Berges:2008mr}. In this regime one may safely choose to study occupancies which, in our case, are derived from equal-time correlation functions in Coulomb gauge.   

For the gauge theory it is a subtle question of how to interpret the most infrared modes beyond the kinetic regime in this way. It has been argued that the infrared cascading solutions observed for scalars~\cite{Berges:2008wm} can be carried over to non-abelian gauge theories~\cite{Carrington:2010sz}. It would be very desirable to devise suitable gauge-invariant measures, in particular, of condensation. This non-trivial task is beyond the scope of the present work. 

This Letter is organized as follows. In section \ref{sec:lattice} we describe the  classical-statistical lattice gauge theory approach, and the results obtained from initial over-population. In section \ref{sec:pert} we compare the numerical results to perturbative computations of scaling exponents in the presence of a time-dependent background field. We conclude in section \ref{sec:conclusion}.

\section{Classical lattice gauge theory simulations}
\label{sec:lattice}

We study the real-time evolution of classical-statistical Yang-Mills theory following closely Ref.~\cite{Berges:2007re}, to which we refer for further technical details. For the simulations we employ the Wilsonian lattice action for SU($2$) gauge theory in Minkowski space-time:
\begin{eqnarray}
S[U] &=& - \beta_0 \sum_{x} \sum_i \left\{ \frac{1}{2 {\rm Tr}
\mathbf{1}} \left( {\rm Tr}\, U_{x,0i} + {\rm Tr}\, U_{x,0i}^{\dagger}
\right) - 1 \right\}
\nonumber\\
&& + \beta_s \sum_{x} \sum_{i<j} \left\{ \frac{1}{2 {\rm
Tr} \mathbf{1}} \left( {\rm Tr}\, U_{x,ij} + {\rm Tr}\, U_{x,ij}^{\dagger}
\right) - 1 \right\} ,
\label{eq:LatticeAction}
\end{eqnarray}
with $x = (x^0, {\bf x})$ and spatial Lorentz indices $i,j = 1,2,3$. It is given in terms of the plaquette variable $U_{x,\mu\nu} \equiv U_{x,\mu} U_{x+\hat\mu,\nu}
U^{\dagger}_{x+\hat\nu,\mu} U^{\dagger}_{x,\nu}$,
where $U_{x,\nu\mu}^{\dagger}=U_{x,\mu\nu}\,$. Here $U_{x,\mu}$ is the
parallel transporter associated with the link from the neighboring
lattice point $x+\hat{\mu}$ to the point $x$ in the direction of
the lattice axis $\mu = 0,1,2,3$. The definitions
$\beta_0 \equiv 2 \gamma {\rm Tr} \mathbf{1}/g_0^2$ and
$\beta_s \equiv 2 {\rm Tr} \mathbf{1}/(g_s^2 \gamma)$
contain the lattice parameter $\gamma \equiv a_s/a_t$, where $a_s$ denotes the spatial and $a_t$ the temporal lattice spacings, and we will consider $g_0 = g_s = g$. 

Varying the action (\ref{eq:LatticeAction}) w.r.t.\ the spatial link variables $U_{x, j}$ yields the classical lattice equations of motion. Variation w.r.t.\ to a temporal link gives the Gauss constraint. We define the gauge fields as
\begin{equation}
g A_i^a(x) \,=\, -\frac{i}{2 a_s} \, {\rm Tr} \left( \sigma^a U_i(x) \right) 
\label{eq:compute-gauge-field}
\end{equation}
where $\sigma^1$, $\sigma^2$ and $\sigma^3$
 denote the three Pauli
matrices. The coupling $g$ can be scaled out of the classical equations of motion and we will set $g = 1$ for the simulations. The
initial time derivatives $\dot{A}_{\mu}^a(x^0=0, {\bf x})$ are set to
zero, which implements the Gauss constraint at
all times. Shown results are from computations
on spatial lattices with $N^3 = 128^3-256^3$ sites.

In order to make contact with discussions in the literature, which are typically formulated in terms of Boltzmann-type equations, we need to extract suitably gauge-fixed distribution functions. These are related to equal-time correlation functions, which are obtained by repeated numerical integration of the classical lattice equations of motion and Monte Carlo sampling of initial conditions~\cite{Berges:2007re}. The dynamics is solved in temporal axial gauge. Though this gauge is very efficient for computational purposes, it involves already on the perturbative level spurious poles in the propagator which makes the extraction of sensible distribution functions or dispersion relations difficult. 
Therefore, we choose to transform the configurations to Coulomb gauge. The gauge transformation to achieve Coulomb gauge is calculated by the overrelaxation method described in Ref.~\cite{Cucchieri:1995pn}. We typically use $10^5$ overrelaxation steps, which achieves a very good convergence for the employed lattice sizes. The Coulomb gauge distribution functions can safely be extracted for momenta where perturbation theory is useful, although it is maybe unclear how to interpret the
most infrared modes in this way. Fortunately, we will see that important lessons can already be learned for not too low momenta such that a characterization of physics in terms of distribution functions is useful.

Distributions will be computed with the help of
time-dependent field momentum modes, which are obtained as $ A^a_i(t,{\bf p})= \int d^3x A_i^a(t,{\bf x}) \exp({i {\bf p \cdot x} })$. We average over color and Lorentz indices for better statistics. 
The occupation number distribution we then define as
\begin{equation}
 n_p(t)  \, = \, \frac{1}{V} \sqrt{ \langle |E(t, {\bf p})|^2 \rangle_{\rm cl} \langle |A(t,{\bf p})|^2 \rangle_{\rm cl} } \, , 
\end{equation}
where $\langle \ldots \rangle_{\rm cl}$ also denotes the classical average over typically five to ten runs and $V$ is the volume. Correspondingly, we consider the ``dispersion''  
\begin{eqnarray} 
 \omega_p(t) \, = \, \sqrt{ \frac{\langle | E(t,{\bf p}) |^2 \rangle_{\rm cl}}{ \langle | A(t,{\bf p}) | ^2 \rangle_{\rm cl}} } \, ,
 \label{dispdef}
\end{eqnarray}
which in Coulomb gauge should only deviate from a free, linear behavior in the regime of large occupancies at low momenta.

\begin{figure}[t]
\includegraphics[scale=1.0]{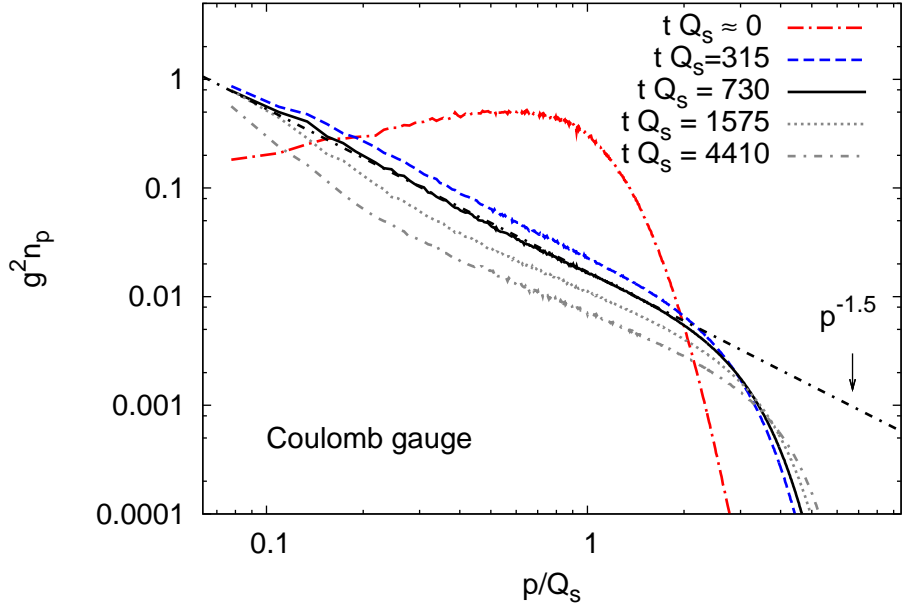}
\caption{Occupation number distribution as a function
of momentum for different times.} 
\label{fig:flatinitial4}
\end{figure}
As mentioned above, we want to study the time-evolution starting from initial over-population. This means that at early times the distribution has a non-perturbatively large occupancy $n_p(Q_s) \sim 1/g^2$ at the characteristic momentum scale $Q_s$. We use Gaussian initial conditions such that all the relevant information about the initial state is contained in the (equal-time) two-point correlation functions. In Fig.~\ref{fig:flatinitial4} we show the occupation number distribution in Coulomb gauge as a function of spatial momentum for different times in units of $Q_s$. The (red) dashed-dotted line shows the approximate initial distribution as it is shortly after the electric field components built up, since we start from $E_i^a = - \dot{A}_i^a = 0$ at $t=0$ to fulfill the Gauss constraint. Relatively quickly we observe the evolution towards a distribution, which for an intermediate time range can be well described by a power-law for momenta $|{\bf p}| \lesssim 2 Q_s$. The shown (blue) dashed curve at $t Q_s = 315$ and the (black) solid one at $t Q_s = 730$ indicate that the power-law behavior is rather well described by the dashed-dotted fit curve $\sim |{\bf p}|^{-3/2}$. 

\begin{figure}[t]
\includegraphics[scale=1.0]{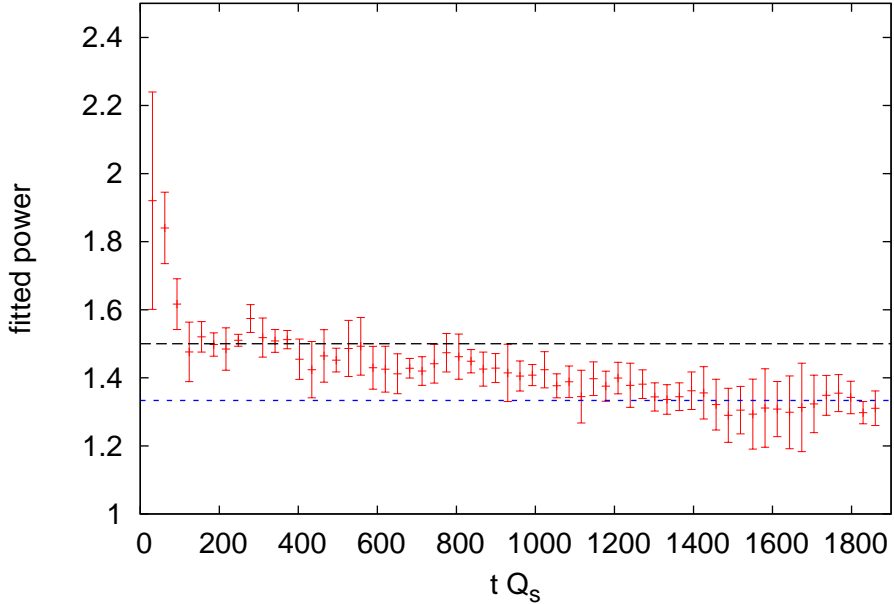}
\caption{The exponent of a fitted power law to the occupation number distribution
displayed in Fig.~\ref{fig:flatinitial4} as a function of time.
The error bars correspond to the fluctuation of the power as the fitting region is varied as explained in the main text.}
\label{fig:fitpower}
\end{figure}
Once this power-law behavior is established, the subsequent evolution becomes rather slow or quasi-stationary. As time proceeds, at some point deviations from a simple power become visible. The (grey) dotted curve for $t Q_s = 1575$ shows already a somewhat steeper behavior at lower momenta and a diminished slope at higher momenta. At $t Q_s = 4410$ the curve is clearly not described by a simple power. 
In order to characterize the transition to and the evolution away from the $\sim |{\bf p}|^{-3/2}$ behavior, we fit a power-law dependence. The resulting exponent as a function of time is shown in Fig.~\ref{fig:flatinitial4}. 
The error bars correspond to the change of the power
 as the fitting region is varied by choosing the lower bound
 from $ (0.3 - 0.6) Q_s $ and the upper bound from $ (0.9 - 1.3) Q_s $.
 The results indicate that initially the system evolves rapidly to $3/2$ (long-dashed curve) and spends a relatively long time around that value, before it slowly evolves to lower values. Of course, this evolution towards lower values is expected since the system will thermalize classically to a distribution $\sim |{\bf p}|^{-1}$ at sufficiently late times (which in general happens rather quickly when the occupied high-momentum modes reach the highest available momenta on the lattice). However, the observation that the intermediate evolution is very slow and stays also a significant time around the value $4/3$ is consistent with earlier investigations starting from initial conditions featuring plasma instabilities~\cite{Berges:2008mr}. This adds to our observation that the observed apparent scaling behavior happens for a wide range of initial conditions. 

\begin{figure}[t]
\includegraphics[scale=1.0]{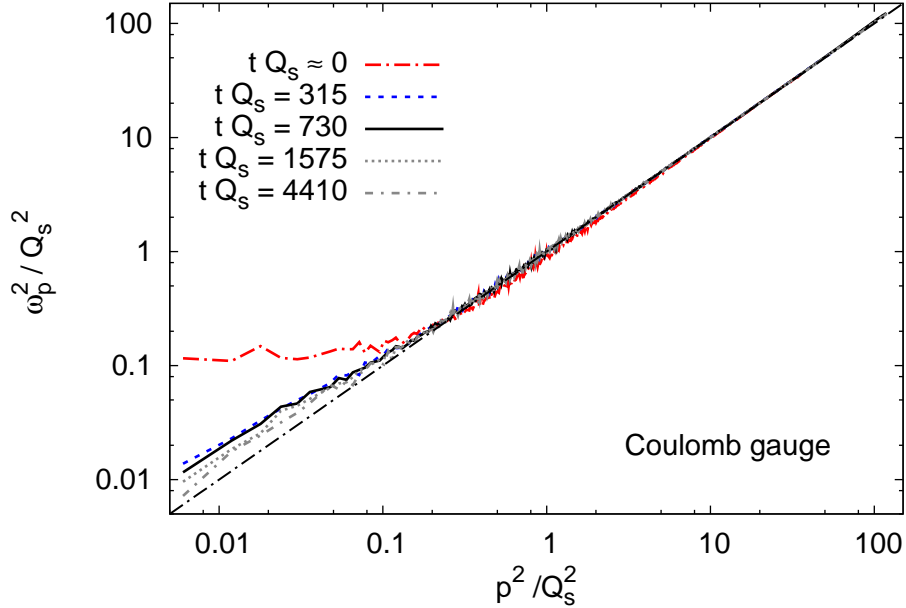}
\caption{Dispersion as a function 
of momentum for different times.}
\label{fig:flatdisp2}
\end{figure}
Before we analyze the power-law behavior in more detail in section \ref{sec:pert} below, we present in Fig.~\ref{fig:flatdisp2} the dispersion. For sufficiently high momenta the considered quantity is always close to the free expression $\omega_p= |{\bf p}|$ as it should. We find that discrepancies from the massless dispersion are present only below $ {\bf p}^2 \lesssim 0.1 Q_s^2 $, with decreasing amplitude as the time grows. Above we explained that the occupation number distribution seems to follows a power-law for intermediate momenta at not too late times. Fig.\ref{fig:flatinitial4} shows that the UV part of the spectrum is very slowly filled up, moving the breakdown of the power-law dependence in the direction of high momenta. In the IR, the breakdown of the power-law dependence is roughly at the same momenta where the dispersion starts to deviate from the free dispersion.

\section{Perturbative scaling exponents}
\label{sec:pert}

For quantum fields $A^a_{\mu}(x)$
one can define two independent connected two-point correlation functions out of equilibrium, which may be associated to the anti-commutator and the commutator,
\begin{eqnarray}
F_{\mu \nu}^{ab}(x,y) & = & \frac{1}{2} \langle \, \{ A^a_{\mu}(x), A^b_{\nu}(y) \} \, \rangle - \mathcal{A}^a_{\mu}(x) \mathcal{A}^b_{\nu}(y) \, , 
\nonumber\\
\rho^{ab}_{\mu \nu}(x,y) & = & i \langle \, [ A^a_{\mu}(x), A^b_{\nu}(y) ] \, \rangle \, ,
\label{eq:def-F-quantum}
\end{eqnarray}
respectively. Here we took into account a possible expectation value or ``background field'' 
\begin{equation}
\mathcal{A}^a_{\mu}(x) \, = \, \langle A^a_{\mu}(x) \rangle \, . 
\label{eq:onepoint}
\end{equation}
Loosely speaking, the spectral function $\rho$ determines which states are available, while the statistical propagator $F$ contains the information about how often a state is occupied. The spectral function is related to the retarded propagator $G_{(R)}$ and the advanced one $G_{(A)}$ as $\rho = G_{(R)} - G_{(A)}$. A tremendous simplification of thermal equilibrium is that the spectral and statistical functions are related by the fluctuation-dissipation relation, which is not assumed here~\cite{Berges:2004yj}. 

\begin{figure}[t]
\centerline{
\includegraphics[scale=0.9]{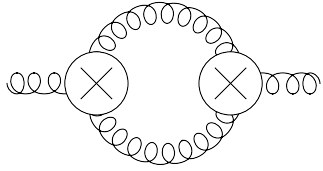}
}
\caption{Gluon part of the one-loop contribution to the self-energy with (2PI) resummed propagator lines. The crossed circles indicate an effective three-vertex in the presence of a background gauge field potential.}
\label{fig:oneloop}
\end{figure}
For instance, the one-loop self-energy correction to two-point correlation functions in the presence of a background gauge potential is diagrammatically given in Fig.~\ref{fig:oneloop}. To avoid problems of secularity in standard perturbation theory, here the lines are meant in the two-particle irreducible (2PI) effective action scheme, where self-energies are expressed in terms of self-consistently dressed propagators \cite{Berges:2004yj}. This includes also the background-field dependence of the propagators. The crossed circles indicate an effective three-gluon vertex $g V_{\mu\nu\gamma}^{abc}$, which is appearing at one loop in the presence of the background gauge field potential (\ref{eq:onepoint})~\cite{Hatta:2011ky}. This effective three-vertex consists of the conventional tree-level vertex and an $\mathcal{A}$-dependent term,
\begin{equation}
V_{\mu\nu\gamma}^{abc} \, = \, V_{0,\mu\nu\gamma}^{abc}+V_{\mathcal{A},\mu\nu\gamma}^{abc} \,\, .
\label{eq:threevertex}
\end{equation}
The standard tree-level part reads in Fourier space
\begin{equation}
V_{0,\mu\nu\gamma}^{abc}(p,q,k) \, = \, f^{abc} \Big( g_{\mu\nu}(p-q)_{\gamma}+g_{\nu\gamma}(q-k)_{\mu}+g_{\gamma\mu}(k-p)_{\nu} \Big) \, , 
\label{eq:V03}
\end{equation}
where $f^{abc}$ are the structure constants of the non-abelian gauge group.
Finally, we will be interested in a situation where the background field has a residual (space-) time dependence. Then the corresponding part of the interaction (\ref{eq:threevertex}) reads in configuration space
\begin{eqnarray}
V_{\mathcal{A},\mu\nu\gamma}^{abc}(x,y,z) & \! = \! &  \left( C_{ac,bd}\, g_{\mu\nu}\, \mathcal{A}_{\gamma}^d(x) + C_{ab,dc}\, g_{\nu\gamma}\, \mathcal{A}_{\mu}^d(x) + C_{ab,cd}\, g_{\gamma\mu}\, \mathcal{A}_{\nu}^d(x) \right) 
\nonumber \\ 
&& g\, \delta^{d+1}(x-y)\, \delta^{d+1}(x-z) 
\label{eq:background}
\end{eqnarray}
with
\begin{eqnarray}
C_{ab,cd}&=&f^{abe}f^{cde}+f^{ade}f^{cbe} \, .
\end{eqnarray}

In order to understand the importance of a non-constant background field, it is instructive to consider first the case of a {\it homogeneous} field $\mathcal{A}_\mu^i(x) = \bar{\mathcal{A}}_\mu^i$, with $\bar{\mathcal{A}}_\mu^i \sim {\mathcal O}(1/g)$. For the case of time and space translation invariant correlators (\ref{eq:def-F-quantum}) we consider their Fourier transform $\tilde{F}(p)$ and $\tilde{\rho}(p)$.\footnote{We introduce a $-i$ in Fourier transforms of the spectral ($\rho$-) and retarded/advanced components, such as $\tilde{\rho}(p) = - i \int {\rm d}^4 x\, e^{ip_\mu x^\mu} \rho(x)$, while $\tilde{F}(p) = \int {\rm d}^4 x\, e^{ip_\mu x^\mu} F(x)$.} Similar to (\ref{eq:def-F-quantum}) we introduce statistical, $\tilde{\Pi}_{(F)}$, and spectral, $\tilde{\Pi}_{(\rho)}$, components of self-energies defined as~\cite{Berges:2004yj}
\begin{eqnarray}
\tilde{\Pi}^{\mu \nu}_{{\rm (}F{\rm )} ab}(p) &=& \tilde{G}_{{(R)} ac}^{-1\, \mu \gamma}(p)\,
\tilde{F}_{\gamma \delta}^{cd}(p)\, \tilde{G}_{{(A)} db}^{-1\, \delta \nu}(p)  \, ,
\nonumber\\
\tilde{\Pi}^{\mu \nu}_{{\rm (}\rho{\rm )} ab}(p) &=& \tilde{G}_{{(R)} ab}^{-1\, \mu \nu}(p)
- \tilde{G}_{{(A)} ab}^{-1\, \mu \nu}(p) \, ,
\label{eq:self-energies}
\end{eqnarray}
where summation over repeated Lorentz and color indices is implied. The translation invariant propagators (\ref{eq:def-F-quantum}) and self-energies (\ref{eq:self-energies}) obey the identity~\cite{Berges:2008sr}
\begin{equation}
\tilde{\Pi}^{\mu \nu}_{{\rm (}F{\rm )} ab}(p)\, \tilde{\rho}_{\nu \mu}^{ba}(p) - \tilde{\Pi}^{\mu \nu}_{{\rm (}\rho{\rm )} ab}(p)\, \tilde{F}_{\nu \mu}^{ba}(p)  \,=\, 0 \, ,
\label{eq:identity}
\end{equation}
which can be directly verified by plugging in the above definitions. Equation (\ref{eq:identity}) is well-known in nonequilibrium physics and will be the starting point for our calculation. In the language of Boltzmann dynamics it states that "gain terms" equal "loss terms" for which stationarity is achieved~\cite{Berges:2004yj}. Thermal equilibrium trivially solves~\eqref{eq:identity}, which we do not consider in the following. Instead, we will look for possible non-thermal scaling solutions.

Decomposing the one-loop self-energy shown in Fig.~\ref{fig:oneloop} into its statistical (real) and spectral (imaginary) part, one obtains
\begin{eqnarray}
\tilde{\Pi}^{\mu \nu}_{{\rm (}F{\rm )}ef}(p;\bar{\mathcal{A}}) &\! = \!& \frac{g^2}{2} \int_{q k} (2\pi)^4\, \delta^{(4)}(p + q + k )\,  V^{\mu \alpha \gamma}_{eac}(p,q,k;\bar{\mathcal{A}}) 
\nonumber\\
&\! \times \!& \left[ \tilde{F}_{\beta\alpha}^{ba}(q)
\tilde{F}_{\delta\gamma}^{dc}(k) + \frac{1}{4} \tilde{\rho}_{\beta\alpha}^{ba}(q) \tilde{\rho}_{\delta\gamma}^{dc}(k) \right]  V^{\nu \beta \delta}_{fbd}(-p,-q,-k;\bar{\mathcal{A}}) \, ,
\nonumber\\
\tilde{\Pi}^{\mu \nu}_{{\rm (}\rho{\rm )}ef}(p;\bar{\mathcal{A}}) &\! = \!& - \frac{g^2}{2} \int_{q k} (2\pi)^4\, \delta^{(4)}(p + q + k )\, V^{\mu \alpha \gamma}_{eac}(p,q,k;\bar{\mathcal{A}}) 
\nonumber\\
&\! \times \!& \left[ \tilde{F}_{\beta\alpha}^{ba}(q) \tilde{\rho}_{\delta\gamma}^{dc}(k)  + \tilde{\rho}_{\beta\alpha}^{ba}(q) \tilde{F}_{\delta\gamma}^{dc}(k) \right]  V^{\nu \beta \delta}_{fbd}(-p,-q,-k;\bar{\mathcal{A}}) 
\label{eq:one-loop}
\end{eqnarray}
with the notation $\int_q \equiv \int d^4 q/(2\pi)^4$ and the symmetry property of the anti-commutator and commutator functions
\begin{eqnarray}
\tilde{F}_{\mu\nu}^{ab}(-p)=\tilde{F}_{\nu\mu}^{ba}(p)\;,\quad \tilde{\rho}_{\mu\nu}^{ab}(-p)=-\tilde{\rho}_{\nu\mu}^{ba}(p)\, .
\end{eqnarray}

Following similar steps as in Refs.~\cite{Berges:2008wm,Berges:2008mr}, scaling solutions can be efficiently identified by integrating (\ref{eq:identity}) over external spatial momentum ${\bf p}$ and suitable scaling transformations for coordinates. In this way the problem can be reduced to simple algebraic conditions for scaling exponents. Non-thermal scaling solutions may be obtained in the classical regime, where expectation values of anti-commutators are much larger than commutators, $F^2 \gg \rho^2$. This is analogous to what is done in the context of weak Kolmogorov wave turbulence using kinetic equations in the regime of sufficiently large occupation numbers~\cite{Zakharov}. In contrast, for lower occupancies of order one, dissipative or quantum corrections will obstruct scaling. 

The stationarity condition then reads
\begin{eqnarray}
&& 0 \, = \, \frac{g^2}{2} \int_{{\bf p}q k} (2\pi)^4\, \delta^{(4)}(p + q + k )\, V^{\mu \alpha \gamma}_{eac}(p,q,k;\bar{\mathcal{A}}) V^{\nu \beta \delta}_{fbd}(-p,-q,-k;\bar{\mathcal{A}})
\nonumber\\
&& \times \left[ \underbrace{\tilde{F}_{\beta\alpha}^{ba}(q) \tilde{F}_{\delta\gamma}^{dc}(k)
\tilde{\rho}_{\nu \mu}^{fe}(p)} + \underbrace{\tilde{F}_{\beta\alpha}^{ba}(q) \tilde{\rho}_{\delta\gamma}^{dc}(k) \tilde{F}_{\nu \mu}^{fe}(p)} + \underbrace{ \tilde{\rho}_{\beta\alpha}^{ba}(q) \tilde{F}_{\delta\gamma}^{dc}(k) \tilde{F}_{\nu \mu}^{fe}(p)}
\right], 
\nonumber\\
&& \qquad \qquad \quad\; {\rm (I)} \; \qquad \qquad \qquad \qquad {\rm (II)} \qquad \qquad \qquad \qquad {\rm (III)}
\label{eq:stat} 
\end{eqnarray}
where $\int_{\bf p} \equiv \int d^3 p/(2\pi)^3$. 
We are looking for scaling solutions which behave as
\begin{equation}
\tilde{F}_{\mu \nu}^{ab}(s p ) = |s|^{- (2 + \kappa )} \tilde{F}_{\mu \nu}^{ab}(p) \quad , \quad \tilde{\rho}_{\mu \nu}^{ab}(s p) = |s|^{-2}\,{\rm sgn(s)}\, \tilde{\rho}_{\mu \nu}^{ab}\left(p \right)  \, 
\label{eq:scaling-assumption}
\end{equation}
under rescaling with the real parameter $s$. This behavior reflects the scaling of the spectral function with the canonical dimension and takes into account a possible occupation number exponent $\kappa$ for the statistical function. This is equivalent to what is obtained from using a kinetic approach. We emphasize that no explicit gauge fixing has to be applied here and the calculation goes through for all gauges which admit scaling solutions of the form (\ref{eq:scaling-assumption}). The scaling properties of the three-gluon vertex (\ref{eq:threevertex}) are different for the standard tree-level part (\ref{eq:V03}) and the background-field part (\ref{eq:background}). To parametrize this behavior in a compact way, we write 
\begin{equation}
V_{\mu\nu\gamma}^{abc}(sp,sq,sk) \, = \, s^v\, V_{\mu\nu\gamma}^{abc}(p,q,k)\;, 
\end{equation}
where $v=1$ for the standard tree-level vertex and $v=0$ for the background field part.

We now map the terms ${\rm (II)}$ and ${\rm (III)}$ in (\ref{eq:stat}) using scaling transformations such that they have the same form as ${\rm (I)}$ up to momentum dependent prefactors. First, we apply for ${\rm (II)}$ the transformation
\begin{equation}
q\, \to \, \frac{p^0}{k^0}\, q \quad , \qquad k\, \to \, \frac{p^0}{k^0}\, p \quad , \qquad p\, \to \, \frac{p^0}{k^0} k \,. 
\end{equation}
The absolute value of the Jacobian for the frequency part of this transformation is $|p^0/k^0|^3$. The same procedure applies to the term ${\rm (III)}$ where the roles of $(k,\gamma,\delta,c,d)$ and $(q,\alpha,\beta,a,b)$ are interchanged. Taking also into account the symmetry of the three-gluon vertex under exchange of combined momentum, color and Lorentz index leads us to 
\begin{eqnarray}
&& 0 \, = \, \frac{g^2}{2} \int_{{\bf p}q k} (2\pi)^4\, \delta^{(4)}(p + q + k )\, V^{\mu \alpha \gamma}_{eac}(p,q,k;\bar{\mathcal{A}}) V^{\nu \beta \delta}_{fbd}(-p,-q,-k;\bar{\mathcal{A}})
\nonumber\\
&& \times \tilde{F}_{\beta\alpha}^{ba}(q) \tilde{F}_{\delta\gamma}^{dc}(k)
\tilde{\rho}_{\nu \mu}^{fe}(p) \left[ 1 + \left|\frac{p^0}{k^0}\right|^{\Delta} \text{sgn}\left(\frac{p^0}{k^0}\right) + \left|\frac{p^0}{q^0}\right|^{\Delta} \text{sgn}\left(\frac{p^0}{q^0}\right)
\right] .
\label{eq:stat2} 
\end{eqnarray}
Here 
\begin{eqnarray}
\Delta & \! = \! & \underbrace{3\cdot3+3}\underbrace{-4}\underbrace{-2(2+\kappa)}\underbrace{-2}    
\underbrace{ +\, v^{(1)} + v^{(2)}}
\label{eq:delta}
\\
&& \mbox{\rm measure} \; \; \mbox{\rm $\delta$'s} \quad \; \, \mbox{ $F F$} \qquad \,
\mbox{$\rho$} \qquad \; \; \mbox{$V V$}
\nonumber
\end{eqnarray}
and $v^{(1)}$, $v^{(2)}$ are the scaling exponents of the respective parts of the two vertex functions appearing in the one-loop self-energy. 

As is well known, the above analysis in the presence of a homogeneous background field is complicated by the fact that to order $g^2$ the individual terms in (\ref{eq:stat}) vanish, since the corresponding processes are kinematically forbidden on-shell, i.e.\ for $p^0 = \pm |{\bf p}|$. Only if the background field has a residual (space-) time dependence, $\mathcal{A}(t)$, the phase space is opened up by the associated frequency contribution of the field and it becomes kinematically allowed. For the above scaling analysis, this amounts to taking \mbox{$v^{(1)} = v^{(2)} = 0$} such that all contributions not involving the background field vanish. After taking this into account, the analysis is analogous to the discussion in scalar field theory where the corresponding dynamics has been analyzed in great detail~\cite{Micha:2004bv,Berges:2012us,Berges:2008wm}.

Therefore, it is also not surprising that the gauge theory in the presence of a time-varying background may exhibit the same scaling exponents as a scalar field theory with quartic self-interaction in the presence of a time-dependent condensate~\cite{Berges:2012us}. 
Indeed, if we set $\Delta = -1$ in (\ref{eq:stat2}) then the $\delta(p^0+q^0+k^0)$ in the integrand of (\ref{eq:stat2}) ensures the vanishing of $(1+k^0/p^0+q^0/p^0)$, using $|k^0/p^0|\, {\rm sgn}(p^0/k^0) = k^0/p^0$ with nonzero $p^0$. From this one can directly read off with (\ref{eq:delta}) the scaling solution
\begin{equation}  
\Delta \, = \, -1 \qquad \Rightarrow \qquad \kappa \, = \, \frac{3}{2} 
\end{equation}
associated to an energy cascade towards higher momenta ~\cite{Zakharov,Micha:2004bv,Berges:2008wm,Berges:2008mr}.

\section{Conclusion}
\label{sec:conclusion}

In this work we have studied the nonequilibrium time evolution of $SU(2)$ gauge theory starting from initial over-population. The classical-statistical simulations reveal a quick evolution towards an approximate scaling behavior, with a subsequent quasi-stationary evolution well described by the scaling exponent $\kappa = 3/2$ for a time duration of hundreds in units of $Q_s$. These lattice results are compared to resummed perturbative estimates at not too low momenta, where the occupancies allow for a kinetic description. Remarkably, the value for the scaling exponent may be understood as arising from the leading ${\cal O}(g^2)$ contribution in the presence of a time-dependent background field. The phenomenon is associated to weak wave turbulence describing an energy cascade towards higher momenta. This particular aspect is very similar to what is observed for scalar theories, where an effective cubic interaction arises because of the presence of a time-dependent Bose condensate. 

Of course, there are important differences between gauge theories and scalar theories. Competing elastic and inelastic scattering processes are expected to prevent a parametrically long time scale for the dominance of a background field dependent contribution in gauge theories, which is consistent with our findings. In particular, the interpretation of the most infrared modes in terms of distribution functions or even the notion of a Bose condensate in gauge theories is non-trivial. The question of whether an intermediate-time behavior is due to a condensate or to very soft modes associated to some slowly varying background-field in time and space may not be too critical after all in view of suggested potential signatures ~\cite{Chiu:2012ij}. For this one also has to consider these questions in classical simulations including the relevant physics of longitudinal expansion in the context of heavy-ion collisions for sufficiently high $Q_s$.  \\[1cm]

\noindent
{\Large \bf Acknowledgments}\\

\noindent
We thank J.-P.~Blaizot, A.~Kurkela, L.~McLerran and G.~Moore for discussions. This work is BMBF (06DA9018) and EMMI supported. We thank the Institute for Nuclear Theory, University of Washington, where this work was completed during the program on {\it Gauge Field Dynamics In and Out of Equilibrium}, March 5 - April 13, 2012.

\end{document}